# Significant enhancement of magnetic shielding effect by using the composite metamaterial composed of mu-near-zero media and ferrite

Xu Chen, Yuqian Wang, Zhiwei Guo,[a)] Xian Wu, Yong Sun, Yunhui Li, Haitao Jiang,[b)] and Hong Chen

*MOE Key Laboratory of Advanced Micro-Structured Materials, School of Physics Science and Engineering, Tongji University, Shanghai 200092, China*

[a)]2014guozhiwei@tongji.edu.cn.
[b)]jiang-haitao@tongji.edu.cn .

The magnetic shield plays an important role in magnetic near-field control. However, the requirements of efficient, ultrathin, lightweight and cheap are still the challenges. Here, we firstly propose a composite metamaterial in which the mu-near-zero media is covered with a ferrite slab. We verify that this structure can enhance the shielding effectiveness in a small area. Furthermore, we optimize the magnetic path by changing the bulk ferrite slab into a patterned slab. In this way, significant shielding effectiveness enhancement can be achieved in a large area. Experimental results show that the maximum shielding effectiveness (SE) of the composite metamaterial with a patterned ferrite is 20.56 dB, which is nearly 19 dB higher than that of a single ferrite slab with the same thickness of the composite metamaterial. The results on the composite metamaterial would be very useful in the applications involving magnetic shielding.

Metamaterials have the ability to control electromagnetic waves that natural materials cannot possess [1], such as negative refraction [2-4], imaging [5, 6], cloaking [7, 8] etc. The unique deep sub-wavelength characteristics enable metamaterials to achieve smaller size and work at lower frequency [9-11]. Nowadays, metamaterials play an important role in controlling near-field electromagnetic waves, which gives optical microcavities or waveguides broader application prospects [12-15]. At present, with the gradual development of near-field technology, a very important issue—magnetic shielding has attracted people's attention. The global QI standard of Wireless Power Consortium requires that the frequency range of wireless power transfer (WPT) is from 100 kHz to 205 kHz [28]. Based on this fact, although metal can shield the magnetic field, the generated eddy currents may cause serious heating effects. Therefore, in order to meet the requirements of the international radio frequency electromagnetic exposure standard [16-18], people usually use ferrite materials or ferrites covered with a metal [20-23]. But this makes the equipment very bulky and expensive. Recently, mu-near-zero (MNZ) media have been utilized in magnetic near-field control. For example, Ref. [24] reported the use of MNZ media as absorbers at GHz frequencies. Ref. [25] proposed the MNZ media for magnetic shielding at MHz frequency, in which the MNZ media is considered as a near-field reflector. However, MNZ medias can only achieve ideal shielding at oblique incidence. Once at normal incidence, there is some magnetic transmission between metamaterial units [25]. And this phenomenon has been observed in the simulated magnetic field results, which directly reduce the shielding effectiveness (SE).

It is worth noting that diamagnetic materials or structure with negative susceptibility have the magnetic shielding effects. High permeability materials can control the magnetic path [26-28]. For example, a shell structure composed of YBCO and ferrite materials, which can be used for magnetic cloaking from dc to 250 kHz [26], and the diamagnetic interface of the active structure also achieves ideal magnetic cloaking [28]. On the other hand, at room temperature, a copper spherical shell structure with equivalent negative susceptibility has similar diamagnetic properties to YBCO in a specific frequency [27]. Similarly, if the MNZ media satisfies the equivalent negative susceptibility ($\chi < 0$), it can be regarded as a diamagnetic interface. More importantly, one can provide an optimal magnetic path for the transmitted magnetic field of the MNZ media by using high permeability materials such as ferrite.

In this work, we show that the MNZ media has equivalent diamagnetic shielding performance at oblique incidence. To improve the SE at normal incidence, we propose a composite metamaterial which combine the MNZ media with a ferrite to optimize the magnetic path and reduce the transmission of magnetic field. Compared with a single MNZ media or a single ferrite material, the significant enhancement of SE is realized by the composite metamaterial composed of the MNZ media and a patterned ferrite. And it has a lighter weight compared with a single ferrite material. Furthermore, in the composite metamaterial, compared with the bulk ferrite slab, the patterned ferrite slab not only has a higher SE, but also greatly reduces the amount of ferrite used. This shows that the patterned ferrite can accurately provide an ideal magnetic path for the transmitted magnetic field of the MNZ media. Therefore, this kind of composite metamaterials would be very useful in the applications of magnetic shielding.

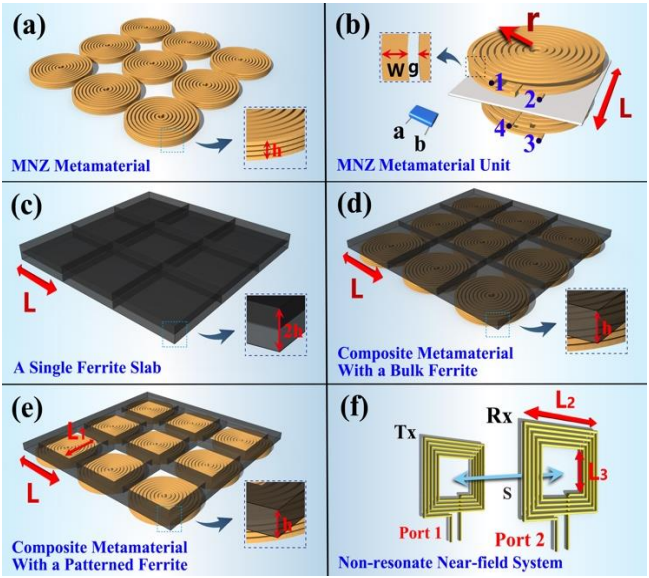

Fig. 1 The schematic diagram of the experimental samples. (a) MNZ media, the total thickness is $h$. (b) MNZ metamaterial unit. The side length of the unit is $L$, and the radius of each single disc is $r$. The width of the copper is $w$, the gap is $g$. The pin a and b of the capacitor is connected to the terminal 2 and 3 by soldering, respectively, and the other terminal 1 is connected to the terminal 4. (c) A single ferrite slab. The side length of each bulk is $L$. The thickness of the slab is $2h$. (d) Composite metamaterials with a bulk ferrite. For each ferrite bulk, the side length is $L$, the thickness is $h$. (e) Composite metamaterials with a patterned ferrite. For patterned ferrite, the outer length is $L$, the inner length $L_1$, and the thickness is $h$. (f) Non-resonant system. The Tx coil is the same as the Rx coil, the outer length is $L_2$, the inner length is $L_3$. The distance between them is $S$.

The scheme of the non-resonant near-field system in the experiment is shown in Fig. 1 (f). The Tx/Rx coils are made of litz wire with a diameter of 6mm and the number of the turns is 10. The outer length of the coil is $L_2 = 150\ mm$, and the inner length $L_3 = 70\ mm$. The two coils are placed symmetrically along the axis, and the distance between each other is $S = 150\ mm$. Then, port 1 and port 2 are connected to the vector network analyzer (Agilent PNA Network Analyzer E5071C). Regarding the design of the metamaterial unit, we choose a pre-pressed $0.1 \times 20$ litz wire as the basic element. As shown in Fig. 1(a), each unit contains four discs with a total thickness h of 2mm. The details are shown in Fig. 1(b). The side length L of the unit is $53\ mm$. The thickness of the outer insulating layer of a single wire is 0.055 mm, thus, the width w of the inner copper is 0.57 mm, and the gap g between the copper is 0.11 mm. Then, the wire is tightly wound to form a disc with 38 turns and a radius of 26 mm. Next, we glue the discs together with polyacrylate adhesive. Here, we chose a CBB non-polar capacitor with a thickness of 2 mm and a capacitance of $8.2\ nF$ as the tuning element. In this case, the thickness of the metamaterial is not enlarged. Finally, the pin a and b of the capacitor is connected to the terminal 2 and 3 by soldering, respectively, and the other terminal 1 is connected to the terminal 4 so that all components form a closed LC resonant circuit (the circuit of the unit is shown in Fig. 1 of Supplementary). However, the diamagnetic effect comes from the mode (tangential wave vector) mismatch between the metamaterial and the environment. When the incident angle is smaller than the critical angle, the magnetic field would transmit through the MNZ medias at normal incidence [25]. To this end, we design a composite metamaterial composed of the MNZ media and a ferrite to improve SE (the parameters of the commercial ferrite are shown in Fig. 2 of supplementary). As shown in Fig. 1(d), the size of the bulk ferrite is $53\ mm \times 53\ mm \times 2\ mm$, where $L = 53\ mm$, $h = 2\ mm$. Therefore, the total side length of the shielding is 159 mm and the total thickness is 4mm. This scheme can only shield a small area close to the back of the bulk ferrite slab (Fig. 3(d)). In order to further achieve a wider range of magnetic shielding, we precisely optimized the path of the transmitted magnetic field between the units by changing the bulk ferrite of MNZ media into a patterned ferrite. In our design, commercially available ferrite materials with the same parameters are used. The patterned ferrite is composed of small ferrite pieces with the size of $20\ mm \times 8\ mm \times 2\ mm$. The outer side length is $L = 53\ mm$, the inner side length is $L_1 = 41\ mm$, and the thickness is $h = 2\ mm$ (Fig. 1(e)). The total side length and total thickness of the two composite metamaterials are the same. Moreover, we also set up a single ferrite slab with the same thickness as the composite metamaterial. The parameters of the ferrite material are the same as above. The total side length of the final shielding body is still 159 mm, and the total thickness is still 4 mm (Fig. 1(c)). Subsequently, the model of the samples is established in the commercial CST Studio Suite simulation software. Generally, it is difficult to model a wire with a multi-strands structure in the software because of a huge amount of calculation. Therefore, a single strand of copper wire is used instead of multiple strands of litz wire (the simulation model of the samples is shown in the Fig. 3 of Supplementary).

According to the effective medium theory, the permeability of the MNZ metamaterials can be written as [19, 25]:

$$\mu = 1 - \frac{F\omega^2}{\omega^2 - \omega_r^2 + j\omega\gamma} \qquad (1)$$

where $F$ is the structural factor, $\omega$ is the frequency, $\omega_r$ is the resonance frequency of the metamaterial, and $\gamma$ is the loss. Since $\chi = \mu - 1$, the susceptibility needs to satisfy [27]:

$$\chi = -\frac{F\omega^2}{\omega^2 - \omega_r^2 + j\omega\gamma} < 0 \qquad (2)$$

where $F, \omega, \omega_r$ and $\gamma$ are positive, so long as $\omega^2 - \omega_r^2 + j\omega\gamma > 0$, that is, when the frequency is greater than resonance and the loss is negligible, the metamaterial is diamagnetic. Then, we obtained the permeability of the metamaterial as shown in Fig. 2(a) according to the effective medium theory [30, 31]. Subsequently, we observe that the structure has a negative susceptibility as shown by the gray area in Fig. 2(b).

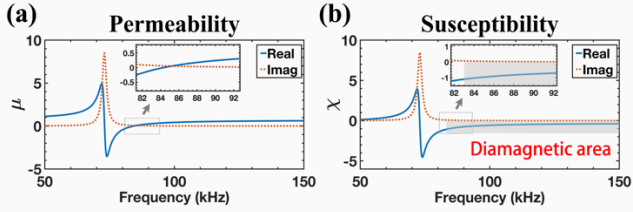

Fig. 2 (a) Permeability of the MNZ media. At 86.23 kHz, $\mu'=0.1$, $\mu''=0.02$ (zoomed view box). (b) Susceptibility of the MNZ media. At 86.23 kHz, $\chi'=-0.9$, $\chi''=0.02$ (zoomed view box). The gray area indicates that the diamagnetic properties start from 83 kHz.

Next, we describe the equivalent diamagnetic mechanism of metamaterials in detail by comparing the relationship between SE and susceptibility. As the frequency increases from 83 kHz, the SE begins to be greater than zero, as shown in Fig. 4(a). Meanwhile, the susceptibility at 83 kHz is negative and the imaginary part is near zero, that is, $\chi'=-1, \chi''=0.05$ (see the black dotted box of Fig. 2(b)), which reflects the diamagnetic property. As the frequency increases to 86.23 kHz, the real part of the susceptibility $\chi'=-0.9$ and the imaginary part $\chi''=0.02$. The loss is further reduced and the SE reaches the maximum value of 17.23 dB, as shown in Fig. 5(a). However, when the frequency is higher than 86.23 kHz, although the imaginary part is negligible, the real part of the susceptibility gradually approaches zero. The diamagnetic performance gradually weakens and the shielding effectiveness gradually decreases.

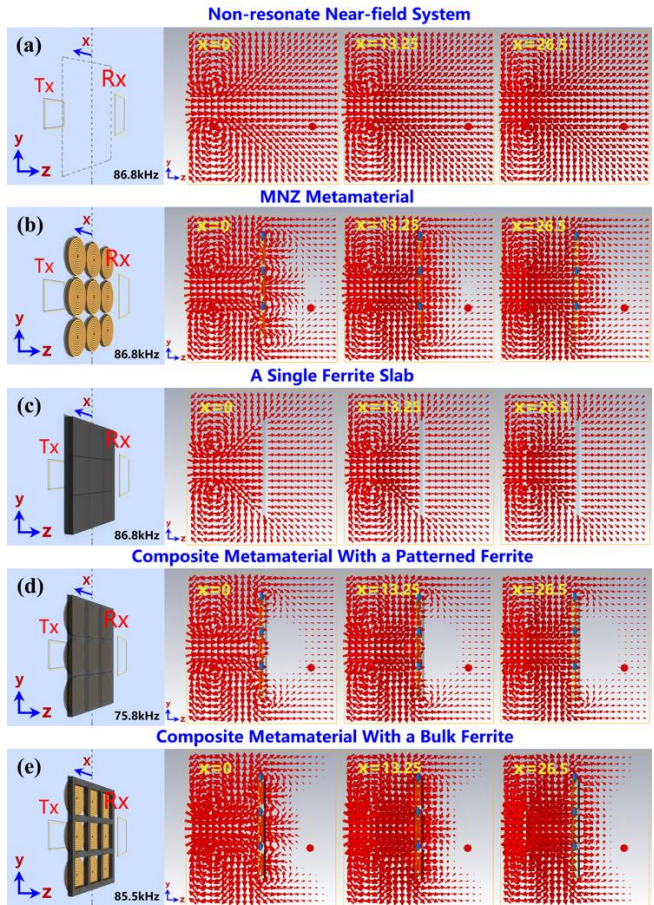

Fig. 3 The scheme of the simulation and the magnetic field results. Since (a) and (c) are non-resonant, the same frequency as (b) was chosen to observe the magnetic field. For other samples, the magnetic field at the frequency of maximum SE is shown. The left side is a scheme of each comparison. Starting from the gray dotted line (x = 0), three yoz slices are selected along the x direction. The slices represent the three positions of the most central unit (center position of the unit: x = 0 mm, 1/4 position of the unit: x = 13.25 mm, boundary position of the unit: x = 26.5 mm).

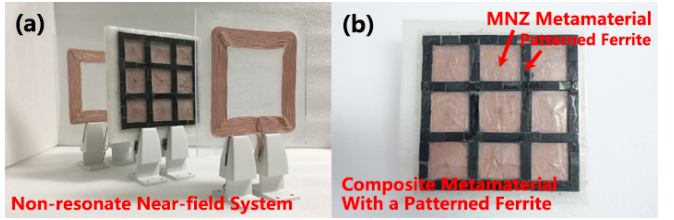

Fig. 4(a) Experimental setup of non-resonant near-field system and composite metamaterial with a patterned ferrite. 4(b) shows the details of composite metamaterial with patterned ferrite.

In order to more intuitively describe the magnetic shielding mechanism of different samples, four samples are positioned between a pair of Tx/Rx coils separated by a distance s and aligned coaxially, respectively. As shown in the schematic diagram on the left in Fig. 3. First, we layer the magnetic field in space, and then select three field slices related to the location of the unit (The center position of the unit $x = 0\ mm$, the 1/4 position of the unit $x = 13.25\ mm$, and the boundary position of the unit $x = 26.5\ mm$). Here, for the non-resonant system and the single ferrite slab, the selected observation frequency is 86.23 kHz. For the other samples, the maximum SE was selected as the observation frequency, respectively. As shown in Fig. 3(a), all the three slices show a strong magnetic field near the Rx coil. When the metamaterial is added to the system, as shown in Fig. 3(b), in the slice with x = 0 mm, the magnetic field near the Rx coil is significantly weaker than before. With the change of the slice position ($x = 13.25 mm$ and $26.5\ mm$), a penetrating magnetic field appears between the units. In order to make up for this shortcoming, considering that ferrite materials have the ability to control the magnetic path [26, 27], we propose a method to construct a composite metamaterial composed of the MNZ media and a bulk ferrite, as shown in Fig. 3(d). The maximum SE of this composite metamaterial is 15.92 dB at 75.8 kHz. On the contrary, compared with the MNZ media, the maximum SE is reduced by 1.3 dB (see Fig. 5(a)). By comparing Fig. 3(b) and Fig. 3(d), after covering the ferrite slab, although the magnetic field is effectively suppressed in a small area near the shielding zone, the severe magnetic flux leakage occurred near the Rx coil. Considering the actual application scenarios, the Tx/Rx coils in the system cannot meet the requirements for accurate placement, and the size of the coils cannot be determined. Therefore, it is more desirable to effectively shield a wider area away from the shielding body. However, using only ferrite for blocking or isolation cannot enhance magnetic shielding. Therefore, to achieve a precise control of the transmitted magnetic field of metamaterials, it is necessary to construct interfaces with a permeability distribution [27, 28]. As a result, we design a patterned ferrite based on the structural characteristics of the metamaterial, as shown in Fig. 3(e). It is a grid structure with square holes. The ferrite part of the grid is aligned at the boundary of the metamaterial unit, and the hole area corresponds to the position of the metamaterial unit. Compared with Fig. 3(b), it can be found that the penetrating magnetic field between the MNZ media units disappears after the patterned ferrite is introduced. Similarly, compared with Fig. 3(d), the magnetic leakage at the position of the Rx coil is perfectly suppressed by adding a patterned ferrite. This achieves a higher degree of freedom and a larger magnetic shielding zone. As shown in Fig. 5(a), the maximum SE of the composite metamaterial with a patterned ferrite at 85.5 kHz is 21.94 dB, which is about 4.7 dB higher than the MNZ media and nearly 6 dB higher than the composite metamaterial with a bulk ferrite. In addition, the SE of a single ferrite slab of the same thickness as the composite metamaterial is always only 4 dB. The field results in Fig. 3(c) show that only a small part of the magnetic field is blocked, and most of the magnetic field transmits.

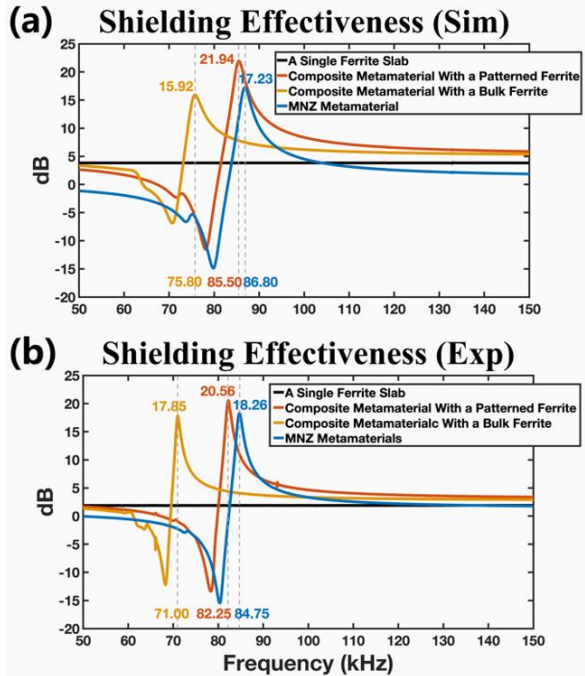

Fig. 5 (a) The shielding effectiveness (SE) obtained by simulation. (b) The shielding effectiveness (SE) obtained by experiment.

Subsequently, the experimental results show that the maximum SE of the composite metamaterial with a patterned ferrite is 20.56 dB at 82.25 kHz. This is even nearly 19 dB higher than the single ferrite slab with the same thickness. Compared with MNZ media, the introduction of patterned ferrite can increase the maximum SE by 2.3 dB. In addition, the maximum SE of the composite metamaterial with a patterned ferrite is 2.7 dB higher than that of the composite metamaterial with a bulk ferrite (see Fig. 5(b)). This fully shows that the patterned ferrite can accurately provide an ideal path for the transmitted magnetic field of the MNZ media. Further, by comparing Fig. 5(a) and Fig. 5(b), it can be seen that the SE of the simulation is slightly wider than the bandwidth obtained from the experiment. Obviously, litz wire makes the metamaterial have lower loss characteristics than the single-turn copper wire.

In conclusion, we show that although the MNZ media has magnetic shielding effect at oblique incidence, it does not work well at normal incidence. To overcome this shortcoming, we use ferrite materials with a high permeability to reduce the magnetic field transmission of MNZ media. Particularly, we find that a patterned ferrite instead of a bulk ferrite can provide an ideal magnetic path for the transmission magnetic field of the MNZ media. Experimental results show that, compared to a single MNZ media or a single ferrite material, the maximum SE of the composite metamaterial with a patterned ferrite is 20.56 dB, which is nearly 2.3 dB higher than that of the MNZ media without ferrite. And this value is even 19 dB higher than that of a single ferrite slab with the same thickness of the composite metamaterial. Particularly, although the patterned ferrite in the composite metamaterial is only 37.6% of the bulk ferrite, the corresponding SE is significantly improved by 2.7 dB. In summary, both simulation and experimental results show that a composite metamaterial with patterned ferrite can significantly improve SE while greatly reducing the weight and volume of the ferrite at the same time. Therefore, this kind of composite metamaterials would have wide applications in the field of magnetic shielding.

This work was supported by the National Key R&D Program of China (Grant No. 2016YFA0301101), the National Natural Science Foundation of China (NSFC; Grant Nos. 12004284, 11774261, and 61621001), the Natural Science Foundation of Shanghai (Grant Nos. 18JC1410900), the China Postdoctoral Science Foundation (Grant Nos. 2019TQ0232 and 2019M661605), and the Shanghai Super Postdoctoral Incentive Program.